\documentstyle[preprint,aps]{revtex}

\begin{document}

\draft

\title{ Order Parameter at the Boundary of a Trapped Bose Gas}

\author{F. Dalfovo, L. Pitaevskii$^{\dagger,*}$
and  S. Stringari}

\address{Dipartimento di Fisica, Universit\`a di Trento, \\
        and Istituto Nazionale di Fisica della Materia,  I-38050
        Povo, Italy } 
\address{$^\dagger$ Department of Physics, TECHNION, Haifa 32000, Israel}
\address{$^*$ Kapitza Institute for Physical Problems, ul. Kosygina 2,
        117334 Moscow }

\date{ April 11, 1996 }
\maketitle

\begin{abstract}
Through a suitable expansion of the Gross-Pitaevskii equation near
the classical turning point, we obtain an explicit solution for the order
parameter at the boundary of a trapped Bose gas interacting 
with repulsive forces. The kinetic energy of the  system, in terms
of the classical radius $R$  and of the harmonic oscillator length
$a_{_{HO}}$,  follows the law  $E_{kin}/N \propto  R^{-2}
[\log (R/a_{_{HO}}) + \hbox{const.}]$,  approaching, for large $R$, the
results obtained by solving numerically  the Gross-Pitaevskii equation. 
The occurrence of a Josephson-type current in  the presence of a double
trap potential is finally discussed. 
\end{abstract}


\narrowtext

The recent experimental realization of Bose-Einstein condensation in 
atomic gases confined in magnetic traps\cite{And95,Bra95,Dav95} is  
stimulating a novel interest in the study of inhomogeneous Bose condensed
systems where the  order parameter exhibits an important spatial
dependence on a macroscopic scale\cite{levico}. 
 
The purpose of the present work is to investigate the behavior of the wave
function of the condensate near the classical turning point, that is, 
at the boundary of the trapped gas.  This region is particularly important
for the determination of the kinetic energy associated with the atoms of
the condensate \cite{BP,DS}.  It is also crucial for the description of
Josephson-type effects taking place in the presence of a barrier in the
confining potential.

The order parameter $\psi({\bf r})$  associated with the ground state  of
a dilute Bose gas obeys the Gross-Pitaevskii equation: 
\begin{equation} 
\left[ - {\hbar^2 \over 2m} \nabla^2 + V_{ext} ({\bf r})
+   {4\pi\hbar^2 a \over m} |\psi({\bf r})|^2 \right] \psi({\bf r}) 
= \mu \psi({\bf r}) \; ,
\label{GP}
\end{equation}
where $V_{ext}$ is the external confining potential, $\mu$ is the chemical
potential and $a$ is the $s$-wave scattering length. The condensate wave
function $\psi({\bf r})$ is normalized to the number $N$ of atoms and is
related to the atomic density through   $\rho({\bf r}) = |\psi({\bf
r})|^2$.  The solution of (\ref{GP}) has been recently found by direct
numerical  integration in the case of both  isotropic \cite{burnett} and
anisotropic traps \cite{DS}.   In the following we will consider
systems interacting with repulsive forces  ($a>0$). When the scattering
length (or the number of atoms in the trap) is sufficiently large, the
solution of Eq.~(\ref{GP}), in the region  where $\mu > V_{ext}({\bf r})$,
takes the simplified Thomas-Fermi form
\begin{equation} 
\psi({\bf r}) = \left[ {m \over 4\pi \hbar^2 a}( \mu -
V_{ext} ({\bf r}) ) \right]^{1/2} \; .
\label{TF}
\end{equation}
Equation (\ref{TF}) is obtained by neglecting the kinetic energy
term $\nabla^2 \psi({\bf r})$ in 
the Schr\"odinger-like equation (\ref{GP})
and provides an accurate description of the exact solution in the interior
of the atomic cloud where the gradients of the wave function
are small, as shown in Fig.~1.

Near the boundary region, where $V_{ext}({\bf r}) \sim  \mu$, the kinetic 
energy term in Eq.~\ref{GP} can not be longer ignored and the
Thomas-Fermi approximation (\ref{TF})  is inadequate. 
This  can have important consequences in the determination of 
relevant physical  quantities. For example if one evaluates   the  kinetic
energy associated with the condensate  
\begin{equation}
E_{kin} = \int \! d{\bf r} \ {\hbar^2 \over 2m} |\nabla \psi({\bf r})|^2
\label{Ekin}
\end{equation}
using Eq.~(\ref{TF}), one finds a logarithmic divergency\cite{BP} which
reveals  that the boundary region must be investigated with higher
accuracy. In the following we will explore the correct behavior of the
order parameter in this region starting from the Gross-Pitaevskii equation
(\ref{GP}). With respect to similar procedures used in the study of the
single-particle Schr\"odinger equation in the presence of an external
field  \cite{LL}, the present method includes explicitly the interatomic 
forces which are responsible for crucial non linear effects in the
equations of motion. 

Let us consider for simplicity a spherical trap\cite{note} and let $R$ be
the  boundary of the system, determined by the equation $\mu=V_{ext}(R)$. 
Near this point one can carry out the expansion
\begin{equation}
V_{ext}(r) - \mu = (r-R) F + o(r-R) 
\label{expansion}
\end{equation}
where $F$ is the modulus of the attractive external  force ${\bf F} = - 
{\bf \nabla} V_{ext}$ evaluated at $r=R$. Close to the
boundary,  where $|r-R| \ll R$, the Gross-Pitaevskii equation takes the
form 
\begin{equation}
 - {\hbar^2 \over 2m} {d^2 \over dr^2} \psi + (r-R) F
\psi +  {4 \pi \hbar^2 a \over m}  \psi^3
= 0\; .
\label{newGP}
\end{equation}
Let us now introduce the dimensionless variable
\begin{equation}
\xi = {(r-R) \over d} 
\label{xi}
\end{equation}
where
\begin{equation}
d = \left({2 m \over \hbar^2} F \right)^{-1/3}
\label{d}
\end{equation}
is a typical thickness of the boundary giving, as we will see later, 
the distance from the classical radius $R$ 
where the Thomas-Fermi approximation starts failing.
Then we introduce the adimensional function $\phi$ defined by
\begin{equation}
\psi({\bf r}) = {1 \over d(8\pi a)^{1/2}}\  \phi(\xi) \; .
\label{phi}
\end{equation}
In terms of $\phi$ the  Gross-Pitaevskii equation (\ref{newGP}) takes the
universal form 
\begin{equation}
\phi^{\prime \prime} - (\xi + \phi^2) \phi = 0 \; .
\label{lev}
\end{equation}
Notice that the non linear term $\phi^3$ arises from the internal 
potential energy in the Gross-Pitaevskii Eq.~(\ref{GP}).  When  $\xi \to
+\infty$ this term can be neglected and  Eq.~(\ref{lev}) takes the 
simpler form $\phi^{\prime \prime} - \xi \phi = 0$ which is the
equation defining the Airy function. The asymptotic 
behavior has then the form  
\begin{equation}
\phi(\xi \to \infty)  \simeq {A \over 2 \xi^{1/4} } \exp
\left( - {2\over 3} \xi^{3/2} \right) \; ,
\label{pinfty}
\end{equation}
where the constant $A$ must be determined by  numerical integration  of 
Eq.~(\ref{lev}).  In the opposite limit $\xi \to -\infty$ one can neglect
the second derivative $\phi^{\prime \prime}$ and the asymptotic  behavior
is given by 
\begin{equation}
\phi(\xi \to -\infty)  \simeq \sqrt{-\xi}
\label{minfty}
\end{equation}
The full behavior of the function $\phi$ is shown in Fig.~2. The  value of
the  constant $A$ is found to be $0.397$.

The solution of Eq.~(\ref{lev}) provides, via Eqs.~(\ref{xi}-\ref{phi}),
the proper structure of the wave function of the condensate near the
classical turning point $R$.  It is worth noting that Eq.~(\ref{lev}) does
not depend on the form of the external potential nor on the size of the
interatomic force. These physical parameters enter the transformations
(\ref{xi}) and (\ref{phi}) which fix, together with the solution of
(\ref{lev}), the  actual behavior of the wave function $\psi$.

Equations (\ref{TF}) and (\ref{phi}) determine the behavior of the 
wave function in two distinct regions of space:  the former in 
the interior of the cloud, the latter in the boundary region. For 
sufficiently large $N$  these two regions are sufficiently
extended to match each other. An example  is shown in  Fig.~1  for
$N=10^5$. 

A third interesting region is the one at large distances beyond the 
boundary $R$ where the system is very dilute and one can ignore the
interaction term in  Eq.~(\ref{GP}). In this region the wave function can
be written in the following way\cite{LL} 
\begin{equation}
\psi(r) = {1\over r}  \left( {\hbar R^2 \over 16\pi d^3a} \right)^{1/2} 
{A \over [(2m(V_{ext}(r) - \mu)]^{1/4}} 
\exp\left(-{\sqrt{2m \over \hbar^2} }\int_R^r [V_{ext}(r^{\prime})-\mu]
^{1/2}dr^{\prime}\right) \; .
\label{asymp}
\end{equation}
The effects of the interatomic interactions enter here
only through the value of the chemical potential.
It is worth noticing that the case $V_{ext}\equiv 0$ would correspond
to the asymptotic behavior of the  order parameter for  saturating systems
in the absence of confining forces as happens, for example, outside the
free surface of superfluid helium \cite{GS}.  The coefficient of
proportionality in (\ref{asymp})  has been fixed in order to obtain the
proper matching with the solution of Eqs.(\ref{phi}-\ref{lev}) in the
region of $r$ such that $R \gg r-R \gg d$ (see Eq.(\ref{pinfty})). 

Let us apply the formalism discussed above  to the 
simplest case of an isotropic harmonic trap:
\begin{equation}
V_{ext}(r) = {1\over 2}m\omega_{_{HO}}^2 r^2 \; .
\label{VHO}
\end{equation}
For $r<R$, the Thomas-Fermi wave function (\ref{TF}) takes the form 
\begin{equation}
\psi_{TF}(r) = \left[{(R^2-r^2) \over 8\pi a_{_{HO}}^4 a}\right]^{1/2}
\label{TFHO}
\end{equation}
where we have used the expression  $\mu = (1/2) m\omega_{_{HO}}^2R^2$
for the chemical potential and introduced the harmonic oscillator 
length  $a_{_{HO}}=(\hbar/m\omega_{_{HO}})^{1/2}$. The radius $R$ is 
fixed by imposing the  normalization of the wave function (\ref{TFHO}) to
the total number of particles: 
\begin{equation}
N = {R^5 \over 15\ a\ a_{_{HO}}^4 }
\label{R}
\end{equation}
and increases very slowly with $N$.

Near the boundary the wave function is instead given  by
Eq.~(\ref{phi})  where the thickness $d$, from Eq.~(\ref{d}),  is 
\begin{equation}
d = \left({a_{_{HO}}^4\over 2 R}\right)^{1/3} \; . 
\label{dHO}
\end{equation}
A similar result for the boundary thickness has been recently 
found by Baym and Pethick (see note 14 in Ref.~\cite{BP}). 
Taking large and negative values of $\xi$ as in Eq.~(\ref{minfty}), 
means moving from the boundary to the interior of the cloud until 
$(R-r) \gg d$. In this region the asymptotic behavior (\ref{minfty}) holds
and one obtains  
\begin{equation} \psi(r) \to 
\left[ { R(R-r) \over 4\pi a  a_{_{HO}}^4}\right]^{1/2}
\label{phiB}
\end{equation}
This exactly coincides with $\psi_{TF}$  given in Eq.~(\ref{TFHO})
provided $(R-r) \ll  R$. In conclusion the wave function in the boundary
region properly matches the Thomas-Fermi wave function (\ref{TFHO})
for values of $r$ satisfying the conditions
\begin{equation}
d \ll (R-r) \ll R \, .
\label{conditions}
\end{equation}
For distances from the boundary less than $d$ the Thomas-Fermi 
approximation (\ref{TFHO}) fails; vice versa, for distances comparable to
the radius $R$, Eq.~(\ref{phi}) becomes inadequate.

Let us apply the above results to the calculation 
of the kinetic energy of the system. The integral (\ref{Ekin}) can be
naturally divided into two parts: 
\begin{equation}
E_{kin} = {4\pi\hbar^2 \over 2m} \left(
\int_0^{R-\epsilon}|\psi^{\prime}(r)|^2r^2dr +
\int_{R-\epsilon}^{+\infty}|\psi^{\prime}(r)|^2r^2dr \right)
\label{2terms}
\end{equation}
where the distance $\epsilon >0$ from the boundary $R$ is chosen 
in such a way that the conditions (\ref{conditions}),  with
$(R-r)=\epsilon$, are satisfied. This permits to evaluate the first term
using the  Thomas-Fermi approximation (\ref{TFHO}), and the second one 
using the solution (\ref{phi}) holding near the boundary.
Clearly the sum of the two terms should not depend on the explicit  value
of  $\epsilon$.

The first integral of Eq.~(\ref{2terms}) is easily evaluated and becomes
\begin{equation}
\int_0^{R-\epsilon}|\psi^{\prime}(r)|^2r^2dr 
= {R^3\over 16\pi a_{_{HO}}^4 a} \left[ 
\log{2R\over \epsilon} -{8\over 3} \right] 
\label{firstterm}
\end{equation}
where we have neglected corrections vanishing as $\epsilon/R$.

For the second contribution, arising from the boundary region,
we instead find the result
\begin{equation}
\int_{R-\epsilon}^{+\infty}|\psi^{\prime}(r)|^2r^2dr =
{R^3 \over 4\pi a_{_{HO}}^4 a}  \int_{-\epsilon/d}^{+\infty}
(\phi^{\prime})^2 d\xi
\label{secondterm}
\end{equation}
where $\phi^{\prime} = d\phi/d\xi$.
If the ratio $\epsilon/d=(R-r)/d$ is sufficiently large (see condition 
(\ref{conditions})) the integral in the right hand side is easily 
calculated and takes the value 
\begin{equation}
\int_{-\epsilon/d}^{+\infty}
(\phi^{\prime})^2 d\xi = {1\over 4} \log{2\epsilon\over d} + C
\label{integral}
\end{equation}
with
\begin{equation}
C = -\int^{+\infty}_{-\infty} \log\left(\sqrt{1+\xi^2}+ \xi\right){d\over
d\xi} \left((\phi^{\prime})^2\sqrt{1+\xi^2}\right)d\xi = 0.176
\label{C}
\end{equation}
In Eq.~(\ref{C}) we have ignored corrections vanishing as $d/\epsilon$.
Collecting the above results and using the explicit expression (\ref{dHO})
for the boundary thickness $d$ in terms of the oscillator  length
$a_{_{HO}}$,  one finally finds the following 
result for the kinetic energy per particle:
\begin{equation}
{E_{kin}\over N} = {5\over 2}{\hbar^2\over mR^2} \left[ \log \left(
{R\over a_{_{HO}}} \right) +  C^{\prime} \right] = {5\over 2}{\hbar^2\over
mR^2} \log \left( {R \over 1.3 a_{_{HO}} } \right) 
\label{kinfinal}
\end{equation}
where $C^{\prime} = (7/4)\log{2}-2+3C$ and we have used 
expression (\ref{R}) for $N$.   Equation (\ref{kinfinal})
provides the correct asymptotic behavior of the kinetic energy
in the limit of large $N$ where $R \gg a_{_{HO}}$. This is 
confirmed by the comparison with the exact value of the kinetic 
energy obtained by solving numerically the Gross-Pitaevskii
equation (\ref{GP}), as shown in Fig.~3. 

We conclude this paper by discussing an interesting application
of the formalism to a Josephson-type effect.
The physical idea is to consider a confining potential 
with two wells separated by a barrier.  When the chemical 
potential in the two traps is different  an oscillating flux 
of atoms is generated.  Let us consider the simplest one-dimensional
problem (extension to 3D will be the object of a future work) and let the
external field $V_{ext}$ consist of two symmetric traps,
trap 1 and trap 2, as shown schematically in Fig.~4.  A difference
between the  chemical potentials $\mu_1$ and $\mu_2$  of the atoms in the
two traps can be achieved, for example, by filling them with a  different
number of atoms. In order to obtain a first analytic result for the 
flux of  atoms generated by the difference in
the chemical potentials  we will assume that 
the barrier between the two wells is high enough. In this 
case the overlap between the wave functions relative to the two
traps, occurs only in the classically forbidden region where interaction
effects can be ignored and one can safely use approximation (\ref{asymp})
for the wave function. Furthermore we will ignore the variation of 
$\mu_1$ and $\mu_2$ generated by the corresponding flux of particles. 
In 1D the factor $1/r$ in the wave
function  (\ref{asymp}) is absent and it is
convenient to take the origin of axes at the symmetry point of the
external potential (see Fig.~4).

The Gross-Pitaevskii equation has two natural solutions in this case.
The first one with chemical potential $\mu_1$, is localized in the trap
$1$. Its  behavior in the classically forbidden region $x>-L_1$ is given by
\begin{equation}
\psi_1(x) = \left({\hbar X_1^2 \over 16\pi d_1^3 a}\right)^{1/2} 
{A \over [(2m(V_{ext}(x) - \mu_1)]^{1/4}} 
\exp \left(-{\sqrt{2m \over
\hbar^2} }\int_{-L_1}^x
[V_{ext}(x^{\prime})-\mu_1]^{1/2} dx^{\prime} \right) \; , 
\label{psi1}
\end{equation} 
where $X_1$ is the distance between the center of trap 1 and the classical
turning point,  $d_1$ is its boundary
thickness [see  Eq.(\ref{d})] and $L_1$ is
the distance  between the classical turning point and the symmetry point
of the external potential (see Fig.~4). 

The second solution with chemical potential $\mu_2$, 
is instead localized in the trap $2$ 
and its behavior in the region $x<L_2$ 
is given by
\begin{equation}
\psi_2(x) = \left( {\hbar X_2^2 \over 16\pi d_2^3 a}\right)^{1/2} 
{A \over [(2m(V_{ext}(x) - \mu_2)]^{1/4}} 
\exp\left(-{\sqrt{2m \over
\hbar^2}} \int_x^{L_2}[V_{ext}(x^{\prime})-\mu_2]^{1/2}
dx^{\prime} \right) \; . 
\label{psi2}
\end{equation}

It is immediate to verify that the linear combination 
\begin{equation}
\psi(x,t) = \psi_1(x) \exp \left( -i{\mu_1 t \over \hbar} \right) +
\psi_2(x) \exp \left(-i{\mu_2 t  \over  \hbar}\right)
\label{1+2}
\end{equation}
is solution of the time dependent Schr\"odinger equation. In fact 
the  wave functions $\psi_1$ and $\psi_2$ significantly overlap only in the
classically forbidden region where non linear
effects due to the interatomic potential are negligible. The current 
density 
\begin{equation}
I = {i\hbar\over 2m}\left(\psi(x,t){\partial \over \partial x}
\psi^*(x,t) -  \psi^*(x,t){\partial \over \partial x}\psi(x,t)\right) 
\label{I}
\end{equation}
associated with the wave function (\ref{1+2})
can be easily calculated  and takes the typical  Josephson form 
\begin{equation}
I=I_0
\sin{(\mu_1-\mu_2)t\over \hbar}
\label{I2}
\end{equation}
with 
$I_0=(\hbar/m)(\psi_1\psi_2^{\prime}-\psi_2\psi_1^{\prime})$.
Using the explicit results (\ref{psi1}) and (\ref{psi2}) for the wave 
functions $\psi_1$ and $\psi_2$ and taking  $\mu_1 \sim \mu_2 = \mu$ and 
$L_1\sim L_2 = L$ in the evaluation of $I_0$,    
we find that the current $I_0$ is uniform in the interval $(-L,+L)$. Its
explicit value is given by the useful result 
\begin{equation}
I_0 = {\hbar A^2  X^2 \over 16 \pi m d^3 a} 
\exp \left(-{\sqrt{2m \over
\hbar^2} } \int_{-L}^{+L}
[V_{ext}(x^{\prime})-\mu]^{1/2} dx^{\prime}\right) 
\label{I0}
\end{equation}

As a consequence of the Josephson current the number of atoms in the
two traps will oscillate in time according to the law\cite{noteJ}
\begin{equation}
{d\over dt}N_1 = -{d\over dt}N_2 = - I_0 \sin{(\mu_1-\mu_2)t \over \hbar}
\, ,
\label{N(t)}
\end{equation}
thereby providing the anticipated result for the flux of particles
through the barrier separating the two traps. 

In conclusion we have obtained an explicit solution of the Gross-Pitaevskii
equation near the classical turning point where the Thomas-Fermi 
approximation turns out to be completely inadequate. Using this 
solution we have been able to derive an analytic expression for 
the kinetic energy of the system holding for large values of 
$N$. We  have finally discussed possible Josephson-type 
oscillations of atoms through the barrier separating  two traps.

\acknowledgements

We thank S.~Vitale for useful discussions.  L.P.  thanks 
the hospitality of the Dipartimento di Fisica at the 
University of Trento.

\begin{figure}
\caption{ Condensate wave function for $10^5$ atoms
of $^{87}$Rb (scattering length $a=5.29 \times 10^{-7}$ cm)
in a spherical  harmonic trap of length
$a_{_{HO}}=1.22 \times 10^{-4}$ cm.  Solid line: numerical solution of the 
Gross-Pitaevskii equation (\protect\ref{GP}). Dot-dashed line:
Thomas-Fermi approximation (\protect\ref{TF}) (indistinguishable from the
solid line in the inner part). Dashed line: surface profile obtained from
the universal equation (\protect\ref{lev}). }
\end{figure}

\begin{figure}
\caption{ Solution of the universal equation (\protect\ref{lev}). 
The two asymptotic limits (\protect\ref{minfty})
(dot-dashed line) and  (\protect\ref{pinfty}) (dashed line) 
are also shown.  } 
\end{figure} 

\begin{figure}
\caption{ Kinetic energy per particle,  in units $\hbar \omega_{HO}$, for
$^{87}$Rb in a spherical harmonic trap as a function of the number of
condensed atoms. Solid line: from the solution of the Gross-Pitaevskii
equation (\protect\ref{GP}). Dashed line: approximation
(\protect\ref{kinfinal}). }
\end{figure}

\begin{figure}
\caption{ Geometry of the double trap for the Josephson effect (see
text).  }
\end{figure} 


\begin{references}

\bibitem{And95} M.H. Anderson, J. R. Ensher, M.R. Matthews, C.E.  
Wieman, and E.A.  Cornell, Science {\bf 269}, 198 (1995).

\bibitem{Bra95} C.C. Bradley, C.A. Sackett, J.J. Tollett, and R.G.
Hulet, Phys. Rev. Lett. {\bf 75}, 1687 (1995).

\bibitem{Dav95}  K.B. Davis, M.-O. Mewes, M.R. Andrews,  N.J. van Druten,
D.S. Durfee,  D.M. Kurn, and W. Ketterle, Phys. Rev. Lett. {\bf 75},
3969 (1995)  

\bibitem{levico} A. Griffin, D.W. Snoke, and S. Stringari,eds, {\it Bose
Einstein Condensation} (Cambridge Univ. Press, Cambridge, 1995). 
This volume collects various review discussions on Bose-Einstein
condensation in different physical systems.

\bibitem{BP} G. Baym and C. Pethick, Phys. Rev. Lett. {\bf 76}, 6
(1996)

\bibitem{DS} F. Dalfovo and S. Stringari, Phys. Rev. A, in press.

\bibitem{burnett} P.A.  Ruprecht, M.J. Holland, K. Burnett, and M.
Edwards, Phys.  Rev.  A {\bf 51}, 4704 (1995).

\bibitem{note} The formalism is  straightforwardly extended to
anisotropic traps. In this case  the distance  $r-R$ entering
Eqs.~(\protect\ref{expansion},\protect\ref{newGP},\protect\ref{xi})  
should be replaced by the quantity $({\bf R}-{\bf r}) \cdot {\bf F}/F$.

\bibitem{LL} L.D. Landau and E.M. Lifshitz, {\it Quantum Mechanics} 
(Pergamon Press, Oxford, 1965)

\bibitem{GS} A. Griffin and S. Stringari, Phys. Rev. Lett. {\bf 76}, 
259 (1996) 

\bibitem{noteJ} The resulting oscillations of $N_1$ and $N_2$ imply a time
dependence in  the difference  between the chemical
potentials in the two traps. Such an effect should be 
small in order to justify the use of Eqs.(\ref{1+2}-\ref{N(t)}). 
This implies  the inequality $I_0 \partial \mu / \partial N \ll 
(\mu_1-\mu_2)^2$, where we have used  $\partial \mu_1/\partial N_1 
\sim  \partial \mu_2/\partial N_2 = \partial \mu/\partial N$.


\end{references}
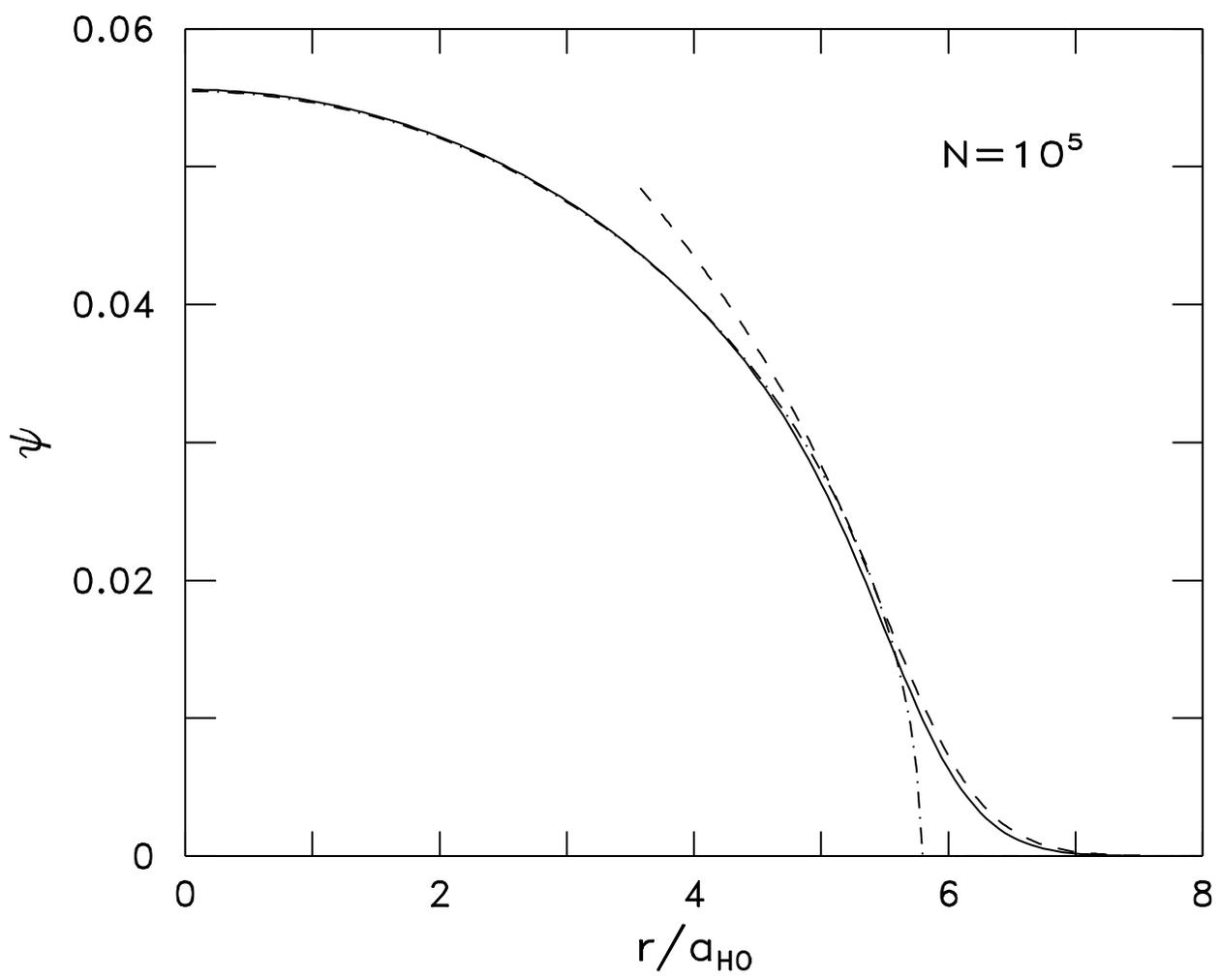
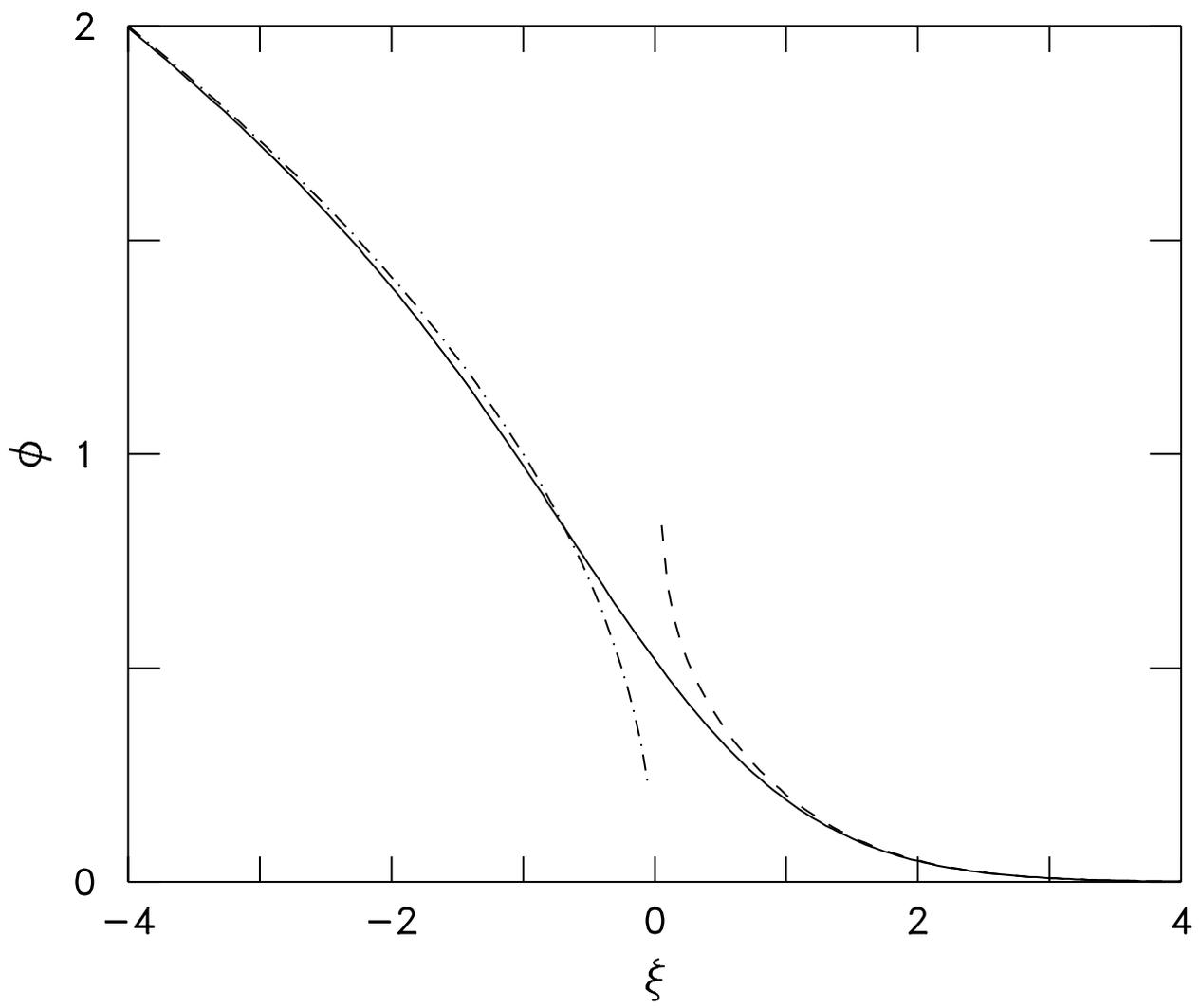
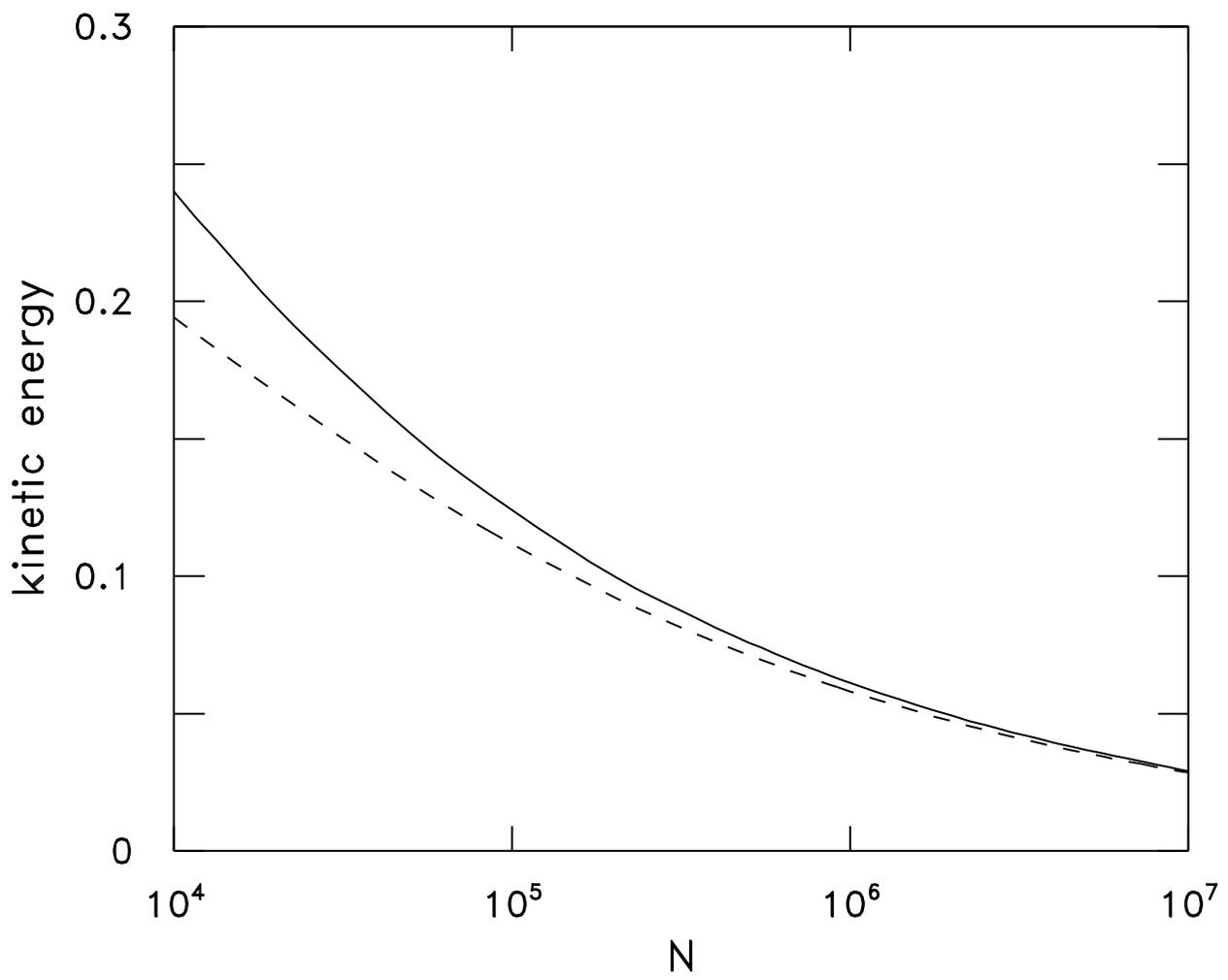
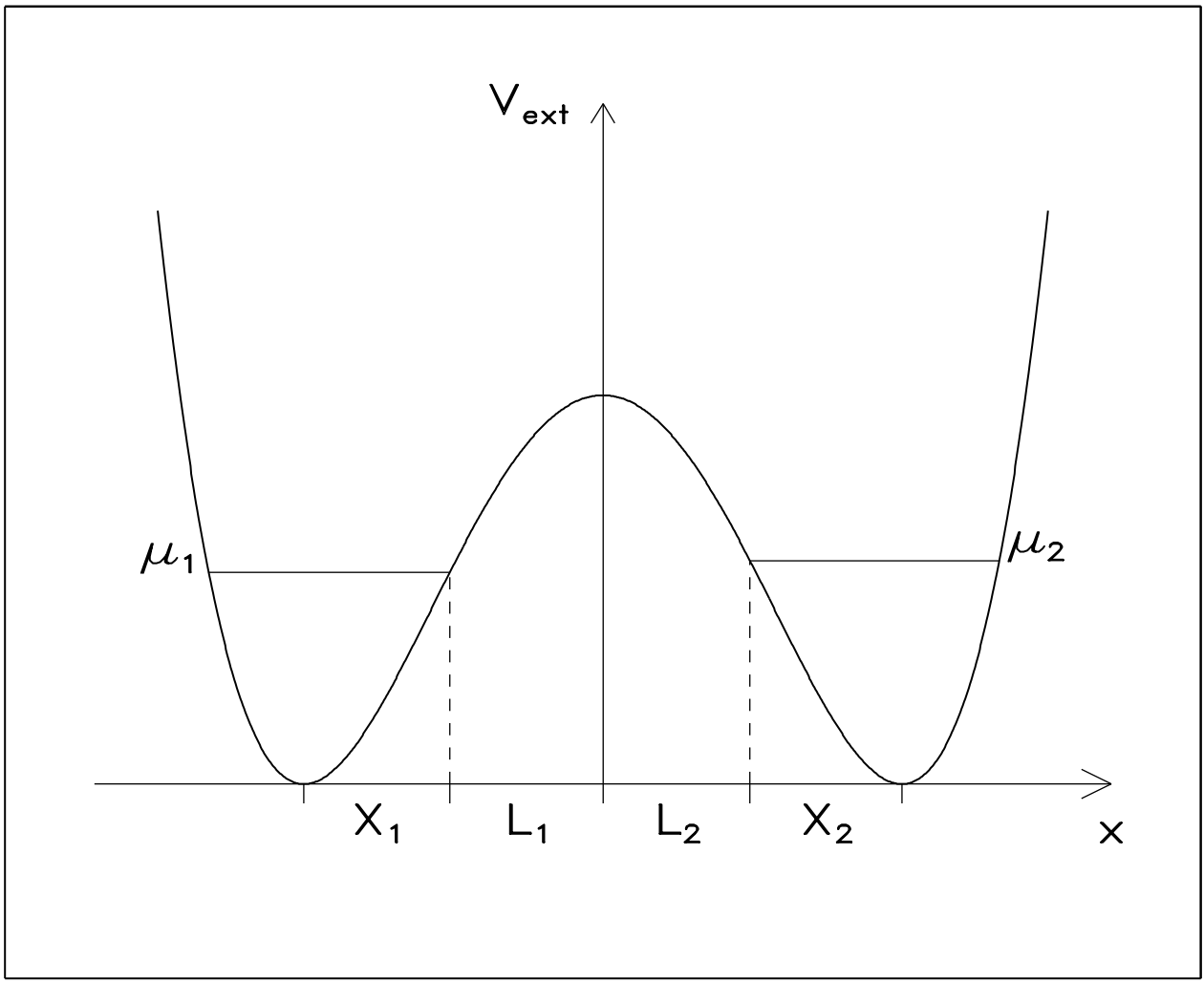
\end{document}